\documentclass[Times,twocolumn,tighten,deluxetable]{aastex7}

\usepackage{graphicx}
\usepackage{rotating}
\usepackage{lineno}
\usepackage{multirow}
\usepackage{enumerate}

\accepted{May 15, 2026}
\submitjournal{ApJ}

\shorttitle{Properties of GS 1354-64 During Its 2025--2026 Outburst}
\shortauthors{Debnath et al.}

\begin{document}

\title{Multimission Observations of GS 1354-64 during the $2025$--$2026$ Outburst: First Results}

\correspondingauthor{Dipak Debnath}
\email{dipakcsp@gmail.com}

\author[0000-0003-1856-5504]{Dipak Debnath}
\affiliation{Institute of Astronomy, National Tsing Hua University, Hsinchu 300044, Taiwan}
\affiliation{Institute of Astronomy Space and Earth Science, P 177, CIT Road, Scheme 7m, Kolkata 700054, India}
\email{dipakcsp@gmail.com}

\author[0000-0002-5617-3117]{Hsiang-Kuang Chang}
\affiliation{Institute of Astronomy, National Tsing Hua University, Hsinchu 300044, Taiwan}
\affiliation{Department of Physics, National Tsing Hua University, Hsinchu 300044, Taiwan}
\email{hkchang@mx.nthu.edu.tw}

\author[0009-0004-5915-3789]{Subham Srimani}
\affiliation{Institute of Astronomy, National Tsing Hua University, Hsinchu 300044, Taiwan}
\email{subhamsrimani2023@gmail.com}

\author[]{Anuj Nandi}
\affiliation{Space Astronomy Group, ISITE Campus, U R Rao Satellite Centre, Bengaluru, 560037, India}
\email{anuj@ursc.gov.in}


\begin{abstract}
The Galactic transient black hole GS 1354-64 recently showed a new outburst, which has been continuously monitored by
{\it MAXI}, {\it NuSTAR} and {\it Insight-HXMT} missions. The ongoing $2025$--$2026$ outburst shows a slow rise with an 
unusually short period ($\sim 3$~days) of X-ray flare of peak flux $\sim 1.4$~Crab, followed by another relatively weak flare of intensity 
$\sim 0.8$ Crab. The source was observed to evolve through ``canonical'' spectral states in a hardness-intensity diagram (HID) during 
rising phase of the outburst, but the subsequent outburst profile during this study, the source did not follow the reverse trend of the HID. 
A rapid evolution of quasi-periodic oscillation (QPO) frequencies ($\sim 0.07-4$ Hz) is observed during hard and intermediate spectral 
states without any signature of QPOs in the soft state. The evolution of the observed low frequency QPOs shows a monotonically increasing 
(rising phase) signature as well as decreasing (decay phase) one, these are further studied with the propagating oscillatory shock model 
to understand the nature of the evolution of the shock wave responsible for the origin of the observed QPOs. 
The broadband energy spectra from {\it NuSTAR} ($3-70$ keV) and {\it Insight-HXMT} ($2-60$ keV) are well described by thermal 
(\textit{diskbb}) and reflection (\textit{relxill}) model components, indicating a strong signature of a relativistic reflection feature. 
Using ``canonical'' observations of a softer state, we found the source to be maximally rotating with $a_k \sim 0.998$ and inclination 
angle to be as $i \sim 63^\circ - 70^\circ$, which are consistent with earlier reports.

\end{abstract}


\section{Introduction}

Compact objects, mainly low mass X-ray binaries, are very fascinating astronomical objects to study in X-rays as they show
variations in their properties in a short duration. Transient black hole X-ray binaries (BH-XRBs) are one of the excellent astronomical 
laboratories to understand the nature of these highly interesting systems \citep{Tanaka95}. Rapid evolution of spectral and 
temporal properties are seen during both rising and declining phases of an outburst of these BH-XRBs 
\citep[see for e.g.,][]{RM06,Belloni05,Nandi12,D15,Sreehari19}. Strong evidence of the correlations between the spectral (photon index, 
disk temperature, election temperature etc.) and temporal (QPO frequency, rms, hardness-ratio etc.) features are observable 
\citep[see for a review,][]{RM06}. The different branches of the Hardness-Intensity Diagram (HID) or the AccRetion Rate Intensity 
Diagram are found to be linked with different spectral states characterized by distinct temporal features, including 
Quasi-periodic Oscillations (QPOs), jets, etc. \citep{Fender04,Belloni05,Homan05,Jana16,Radhika16,Choudhury25}. Generally, four spectral states: 
low-hard (LHS), hard-intermediate (HIMS), soft-intermediate (SIMS) and high-soft (HSS) are observed during a classical or 
type-I outburst of a transient BH candidate, while softer states are found to be missing during a harder or type-II outburst 
\citep{D17}. These spectral states also show a hysteresis loop with LHS as a start and end phase of the outburst 
\citep{Homan05,Nandi12,D13,Sreehari21}.

Transient BH-XRBs generally remain in low luminosity \citep[$L_X \sim 10^{30-33}~erg~s~^{-1}$;][]{Tetarenko16} quiescence phases, and
occasionally become active for a period of weeks to months. During these active or outbursting phases, the sources become extremely 
luminous \citep[$L_X \sim 10^{37-38}~erg~s~^{-1}$;][]{Tanaka96}. It is generally believed that an outburst is triggered due to a 
sudden rise of viscosity at the outer edge of the accretion disk \citep{Ebisawa96}. Recently \citet{C19} has introduced the concept 
of accumulation of mass from the companion donor star at a temporary reservoir, namely the pile-up radius ($X_p$), located at a far 
distance between the BH and Lagrange point L1. In a quiescence or accumulation phase, viscosity as well as instability increases 
at $X_p$ with the increase of the accumulated matter. The onset of an outburst occurs due to the rapid inflow of accumulated 
matter toward the BH when the viscosity exceeds a critical threshold, which depends on $X_p$. 
The triggering mechanism of type-I and type-II outbursts has quite successfully explained 
different types (time period, intensity, peak flux etc.) of outbursts in three recurring transient BH candidates: H~1743-322 \citep{C19}, 
GX~339-4 \citep{Bhowmick21}, 4U~1630-47 \citep{Chatterjee22}. 

The electromagnetic radiations that we observe in multi-wavelength bands, are emitted from the accretion disk formed around the compact 
object. The continuum part of the energy spectrum of a BH-XRB, mainly consists of two components: one multi-color thermal disk 
black body emission from the accretion disk \citep{Mitsuda84, Makishima86a} and another non-thermal power-law tail, which originated 
from the Corona or Compton cloud due to inverse-Comptonization of disk photons \citep{ST80,ST85}. Several models have been proposed to 
explain the nature and origin of this `Compton cloud', such as a magnetic corona \citep{Galeev79} and a hot gas corona above the disk 
\citep{Haardt93,Zdziarski03,Done07}. In the context of two component advective flow model \citep{CT95}, accretion disk contains two different types 
of flows. First one is high viscous, high angular momentum, geometrically thin and optically thick Keplerian disk matter, which forms a standard 
disk at the equatorial plane \citep{SS73} and acts as the source of observed thermal black body photons. The other flow component 
is a low-viscous, low-angular-momentum, optically thin, and geometrically thick sub-Keplerian halo (flanked above and below the Keplerian 
disk), which forms a `hot' puffed-up region (i.e., the corona) due to a shock transition \citep{C96a,C96b}. 
This corona acts as a Compton cloud where thermal disk black body photons from the Keplerian disk become energetic hard photons via multiple 
inverse-Compton scattering \citep{ST80,ST85}. Conversely, hard photons might be generated through upscattering of soft photons via synchrotron processes 
\citep[e.g.,][]{Wardzinski00,Veledina11}. Furthermore, a third component may be present in the BH spectrum, characterized by the reflection 
of hard photons from the disk due to Compton down-scattering and atomic absorption or re-emission 
\citep{Lightman88,Fabian89,Brenneman06,Bambi21,Garcia10}. Occasionally a prominent feature of characteristic line emission from various 
elements (e.g., the $6.5$~keV Fe $K_\alpha$ line) is observed in BH spectra \citep{Makishima86b,Fabian89}. Moreover, the observed skewed 
Fe K$_\alpha$ line and reflection signatures are considered to be strong relativistic effects induced by the spin of the black hole 
\citep{Garcia10,Reynolds14}.

Low frequency ($0.01-30$~Hz) QPOs (LFQPOs) are commonly observed in hard and intermediate spectral states of BH-XRBs \citep[see,][]{RM06}. 
Depending upon the nature (fundamental frequency, rms, Q-value, noise, lag, etc.), QPOs are classified into three main types: A, B, C 
\citep{Casella05}. In the LHS and HIMS of both rising and declining phases LFQPOs (type-C) are found to evolve monotonically, while in the SIMS 
QPOs (type-A or -B) are observed to the sporadically on and off without any detection in HSS. Many model have been proposed to explain origin 
of these QPOs, for example, magneto-acoustic waves \citep{Titarchuk98}, spiral density waves \citep{Varniere02}, Lense-Thirring precession \citep{Stella99}, 
shock oscillation model \citep{MSC96,RCM97,C15}. In the past, attempts also have been made to understand the evolution of QPO frequencies 
observed in various BH-XRBs \citep[][and references therein]{Vignarca03,Titarchuk09, Stiele13}. However, according to the shock oscillation 
model, oscillation of the shock is considered as the origin of different types of QPOs. Type-C QPOs occur due to the resonance oscillation of 
the shock (when cooling and in-fall time scales roughly matches), while type-B QPOs originate from either non-satisfaction of the Rankine-Hugoniot 
conditions (necessary for stable shock formation) or a weakly resonating corona. The possible origin of the broader type-A QPOs might be due to 
the weak oscillations of the shockless centrifugal barrier \citep[see][]{C15}. Furthermore, time varying form of the shock oscillation model, namely 
propagating oscillatory shock (POS) model is found to be quite successful for understanding of monotonic evolution of the QPO 
frequencies during the rising and declining phases of the outbursts of many transient BH-XRBs 
\citep[see for example,][]{C08,Nandi12,D13,D25}. Due to the monotonic rise in the QPO frequency, inward shock motion is 
observed during the rising phase of the outburst, while the reverse scenario is observed during the declining phase, as the 
QPO frequency is found to decrease monotonically.

GS~1354$-$64 (= BW~Cir) is one of the dynamically confirmed stellar-mass BH-XRBs. It was discovered by the \textit{Ginga} satellite 
in 1987 \citep{Swinbanks87,Makino87}. Two earlier black hole sources, namely Cen~X$-$2 \citep{Francey71} and MX~1353$-$64 
\citep{Markert79}, were also reported from the same sky location as GS~1354$-$64. According to \citet{Kitamoto90}, these 
sources are likely to be the same object. Dynamical studies later confirmed GS~1354$-$64 as a black hole with a mass of 
$M_{\rm BH}=7.9\pm0.5~M_\odot$ and an orbital period of $P_{\rm orb}=2.54$~days \citep{Casares09}. \citet{Casares04} identified the 
binary companion as a G$0$–$5$~III star with a mass of $1.1\pm0.1~M_\odot$. Furthermore, \citet{Casares09} constrained the distance and 
inclination angle of GS~1354$-$64 to be $\geq 25$~kpc and $\sim75^\circ$, respectively. From a model-dependent study of the 
reflection-dominated hard-state spectrum during the 2015 outburst, \citet{El-Batal16} estimated the black hole spin and inclination 
angle to be $a_k=0.98$ and $i=75^\circ$, respectively. Using HEASARC column density tool, 
\citet{El-Batal16} adopted a hydrogen column density of $N_{\rm H} = 7 \times 10^{21}~{\rm cm^{-2}}$ for their spectral analysis. 
A similar range of $N_{\rm H} \sim (5$--$9)\times10^{21}~{\rm cm^{-2}}$ is also reported by \citet{Kalberla05} from the Galactic 
H\textsc{I} survey within a $1^\circ$ radius of the location of GS~1354$-$64.

Recently, on 2025 December 25 (MJD~61034), \textit{MAXI}/GSC detected a new outburst of the source \citep{Negoro25}. This detection was 
subsequently confirmed by several missions across multiple wavelengths, ranging from radio and optical to X-rays. \citet{Adegoke26} 
reported the detection of relativistic reflection features based on a preliminary analysis of \textit{NuSTAR} data.
Based on the evolution of the \textit{MAXI}/GSC count rates in the soft and hard energy bands, \citet{Negoro26} reported a transition 
in the spectral nature of the source from the hard to the soft state on 2026 January 19 (MJD~61059). During the outburst, 
a broad signature of Fe emission line with \texttt{XRISM} \citep{Liu26} and detection of polarization with \texttt{IXPE} \citep{Ravi26} are 
also reported.

In this {\it paper}, we present the first results on the spectral and temporal properties of GS~1354$-$64 during its 2025–26 
non-conventional outburst. We analyze archival data from multiple satellites, namely \textit{MAXI}, \textit{NuSTAR}, and 
\textit{Insight-HXMT}, covering the period from the initial phase of the outburst up to 2026 February 19. The paper is organized as 
follows: \S2 describes the observations, data reduction, and analysis methods. In \S3, we present the results of our detailed spectral 
and timing analyses, while \S4 discusses our findings and presents the conclusions.

\section{Observation and Data Analysis}

We study the recent outburst of the Galactic transient BH GS~1354$-$64 using on-demand data from {\it MAXI}/GSC and archival data from 
the {\it NuSTAR}/FPMA and {\it Insight}-HXMT/LE, ME, and HE instruments. We follow standard data reduction procedures for both
{\it NuSTAR}\footnote{\url{https://heasarc.gsfc.nasa.gov/docs/nustar/analysis}} and
{\it Insight}-HXMT\footnote{\url{http://hxmten.ihep.ac.cn/analysis.jhtml}} satellites. The {\it NuSTAR} data are extracted using the 
online platform SciServer\footnote{\url{https://www.sciserver.org}}, employing the latest version (v6.36) of the HEASoft package provided 
by HEASARC. The {\it Insight}-HXMT data analysis is performed locally using the satellite software package within HEASoft (v6.34) after 
downloading the archival data.

\subsection{MAXI}
To study the outburst profile and hardness ratio evolution between 2025 December~25 (MJD~61034) and 2026 February~19 (MJD~61090),
{\it MAXI}/GSC daily-averaged fluxes in the soft ($2$--$6$~keV) X-ray (SXR) and hard ($6$--$20$~keV) X-ray (HXR) bands are downloaded 
from the MAXI RIKEN on-demand website\footnote{\url{https://maxi.riken.jp/mxondem}}. The ratio of the HXR to SXR flux is defined as the 
hardness ratio (HR), while the $2$--$20$~keV flux is referred to as the total X-ray (TXR) flux.

\subsection{NuSTAR}
The {\it NuSTAR} \citep{Harrison13} TOO observations of 2026 Jan 13 (ObsID 91201302002), Jan 21 (ObsID 91201302004), 
Jan 24 (91201302006), Jan 30 (ObsID 91202308002), Feb 11 (91202311002) and Feb 11 (91202311004) are reduced using the 
{\fontfamily{qcr}\selectfont NuSTARDAS} software (v2.1.4). Cleaned event files are produced with the {\fontfamily{qcr}\selectfont nupipeline} 
task using the latest calibration files. Source spectra and light curves (in $10$~s, $1$~s, $0.01$~s bins) are extracted from a circular 
region of $50''$ radius centered at the source coordinates, while corresponding background files are extracted from an annular region 
($180''-240''$) defined using DS9. The standard {\fontfamily{qcr}\selectfont nuproducts} task is used to generate source spectra, 
auxiliary response files, and response matrix files. To improve signal-to-noise ratio, we further rebinned spectra using 
{\fontfamily{qcr}\selectfont GRPPHA} task to have at least $100$ counts per bin.

\subsection{Insight-HXMT}
We study publicly available archival data of $37$ observations of {\it Insight}-HXMT \citep{Zhang14, Zhang20} in between 2026 Jan 18 
(MJD 61058.18) to Feb 08 (MJD 61079.34) are extracted using latest calibration files (v2.08) to study detailed temporal and spectral 
properties of the source. We generate $1$~s and $0.01$~s time binned light curves in different energy bands of LE ($2-4$~keV, $4-10$~keV,
$2-10$~keV), ME ($8-30$~keV, $10-30$~keV), and HE ($27-150$~keV). For the spectral analysis, we used data in default energy bands data, 
which covers maximum energy band of LE, ME, HE instruments. The data extraction for generating light curves and spectral files was 
performed using {\tt hpipeline} pipeline within the HEASoft software package (v.6.34) from HEASARC. To improve signal-to-noise ratio, 
we also rebinned spectra to ensure at least $50$ counts per bin using the {\fontfamily{qcr}\selectfont GRPPHA} task.

\subsection{Data Analysis}
To search for the key temporal feature, namely QPOs, we generated Fast Fourier transform power density spectra (PDS) using $0.01$~s 
time-binned light curves from the {\it NuSTAR} and {\it Insight-HXMT} satellites. We used the 
{\fontfamily{qcr}\selectfont XRONOS} task {\fontfamily{qcr}\selectfont powspec} (part of the HEASoft software package) with 
{\fontfamily{qcr}\selectfont norm=2}, which produces PDS in units of squared fractional rms variability without white-noise subtraction.
To determine the QPO parameters: centroid frequency ($\nu_{QPO}$), full width at half maximum (FWHM), and normalized power, the PDS were fitted using a 
combination of Lorentzian and constant models.

The broadband {\it Insight-HXMT} spectrum in the $2$–$60$~keV energy range after combining data from LE ($2$–$8$~keV), ME ($8$–$30$~keV), 
and HE ($27$–$60$~keV), was fitted with the following models --
M1: {\fontfamily{qcr}\selectfont constant$\times$tbabs(diskbb+powerlaw)},
M2: {\fontfamily{qcr}\selectfont constant$\times$tbabs(thcomp$\otimes$diskbb)}, and
M3: {\fontfamily{qcr}\selectfont constant$\times$tbabs(diskbb+relxill)}. Similarly, the $3$–$70$~keV {\it NuSTAR}/FPMA spectrum was 
fitted using the same three model combinations. Here, to obtain statistically acceptable fits, an additional Gaussian component was 
required in model combination M1 and M2. 
For spectral fit, we used fixed value of the hydrogen column density $N_{\rm H} = 7 \times 10^{21}~{\rm cm^{-2}}$ for absorption 
model \texttt{tbabs}.

\begin{figure}
  \centering
    \includegraphics[angle=0,width=8.8cm,keepaspectratio=true]{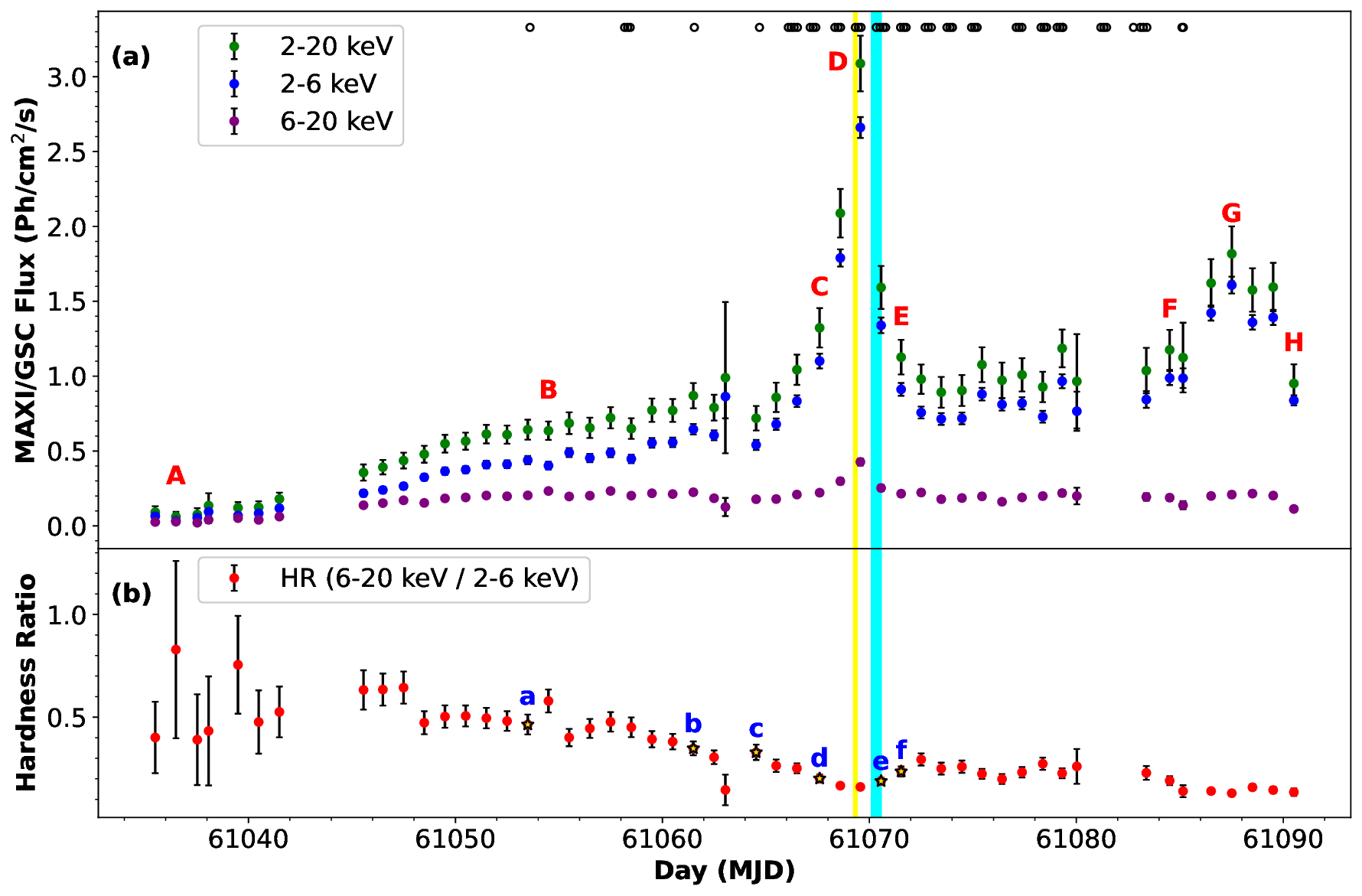}
    \caption{Variation of (a) \textit{MAXI}/GSC fluxes in the SXR ($2$–$6$~keV), HXR ($6$–$20$~keV), and TXR ($2$–$20$~keV) energy bands, 
    and (b) the hardness ratio (HR), defined as the ratio of HXR to SXR fluxes. The yellow and cyan shaded regions indicate the observation 
    periods of the \textit{Insight-HXMT} and \textit{NuSTAR} satellites, respectively, for which detailed spectral analyses are performed. 
    The points A–H in the upper panel mark different stages of the HID (see Fig.~\ref{fig_HID}), while the points labeled (lower panel) a–f 
    denote the MJDs corresponding to different stages of the QPO evolution (see Fig.~\ref{fig_pos}).}
\label{fig_lc}
\end{figure}

\section{Results}

The detailed temporal and spectral properties during the initial phase of the outburst of the Galactic transient BH GS~1354$-$64 are 
studied using data from three satellite instruments, namely {\it MAXI}/GSC, {\it NuSTAR}/FPMA, and {\it Insight}-HXMT/LE, ME, and HE. 
The {\it MAXI} data are used to investigate the daily evolution of the outburst profile and hardness ratios, while the {\it NuSTAR} 
and {\it Insight}-HXMT data are used for detailed temporal and spectral studies of the source. In the following subsections, 
we present our analysis results based on observations from these multimission satellite instruments.

\begin{figure}
  \centering
    \includegraphics[angle=0,width=8.5cm,keepaspectratio=true]{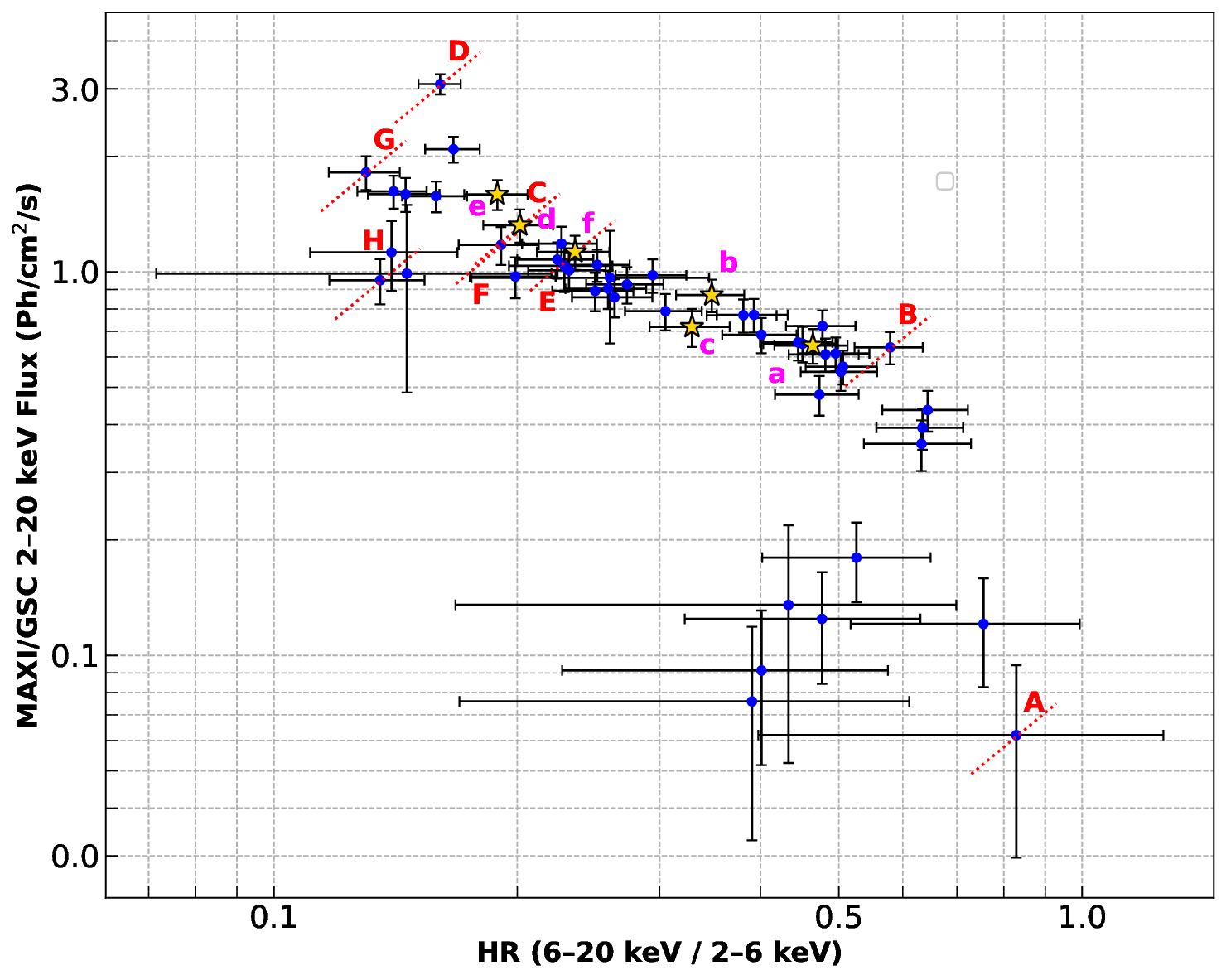}
	\caption{The MAXI/GSC hardness–intensity diagram (HID) from 2025 December 27 (MJD~61036) to 2026 February 19 (MJD~61090) is shown. 
    The dashed-line points A, E, and D indicate the start, end, and peak of the initial phase of the outburst, respectively, whereas 
    the dotted-line points B and C mark the onset of strong inflow in the SXR, triggering an X-ray flare and the end of the flare, respectively. 
    The star-shaped points {\rm a–f} denote different evolutionary phases of the QPO, as indicated in Fig.~\ref{fig_pos}. 
    Both X- and Y-axes are shown on logarithmic scales.}
\label{fig_HID}
\end{figure}

\subsection{Outburst Profile and Hardness-Intensity Diagram}

{\it MAXI} monitored the new outburst of GS~1354$-$64 on an approximately daily basis, starting from its detection on 2025 December 25 
(MJD~61034). In Fig.~\ref{fig_lc}(a), we present the evolution of the daily averaged fluxes in the soft ($2$–$6$~keV), hard ($6$–$20$~keV), 
and total ($2$–$20$~keV) X-ray bands of the {\it MAXI}/GSC from 2025 December 27 (MJD~61036) to 2026 February 19 (MJD~61090). These three 
energy bands are hereafter referred to as SXR, HXR, and TXR, respectively. The points {\rm A–H} mark the start, end, or transitions between 
important evolutionary phases of the 2025–26 outburst. The yellow and cyan shaded regions indicate the soft-state observation periods of 
{\it Insight}-HXMT and {\it NuSTAR}, respectively, for which detailed spectral analyses were performed to estimate the spin and inclination 
angle of the source. In Fig.~\ref{fig_lc}(b), we show the evolution of the hardness ratio (HR), defined as the ratio of the HXR to SXR fluxes. 
The source exhibits strong low-frequency QPOs and their temporal evolution. The points {\rm a–f} mark key phases of the QPO frequency 
evolution, the details of which are discussed in the following subsection.

\begin{figure*}
  \centering
    \includegraphics[angle=0,width=8.8cm,keepaspectratio=true]{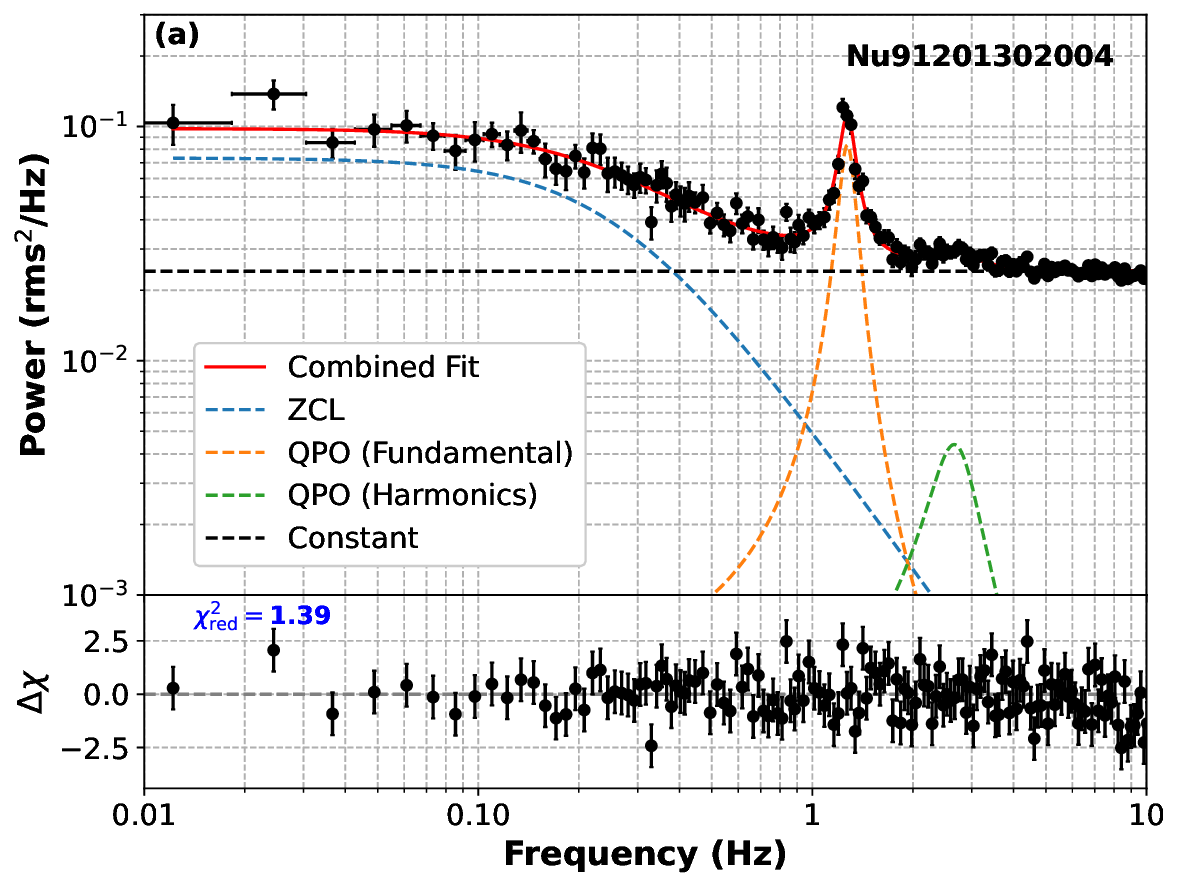}\hskip 0.02cm
    \includegraphics[angle=0,width=8.8cm,keepaspectratio=true]{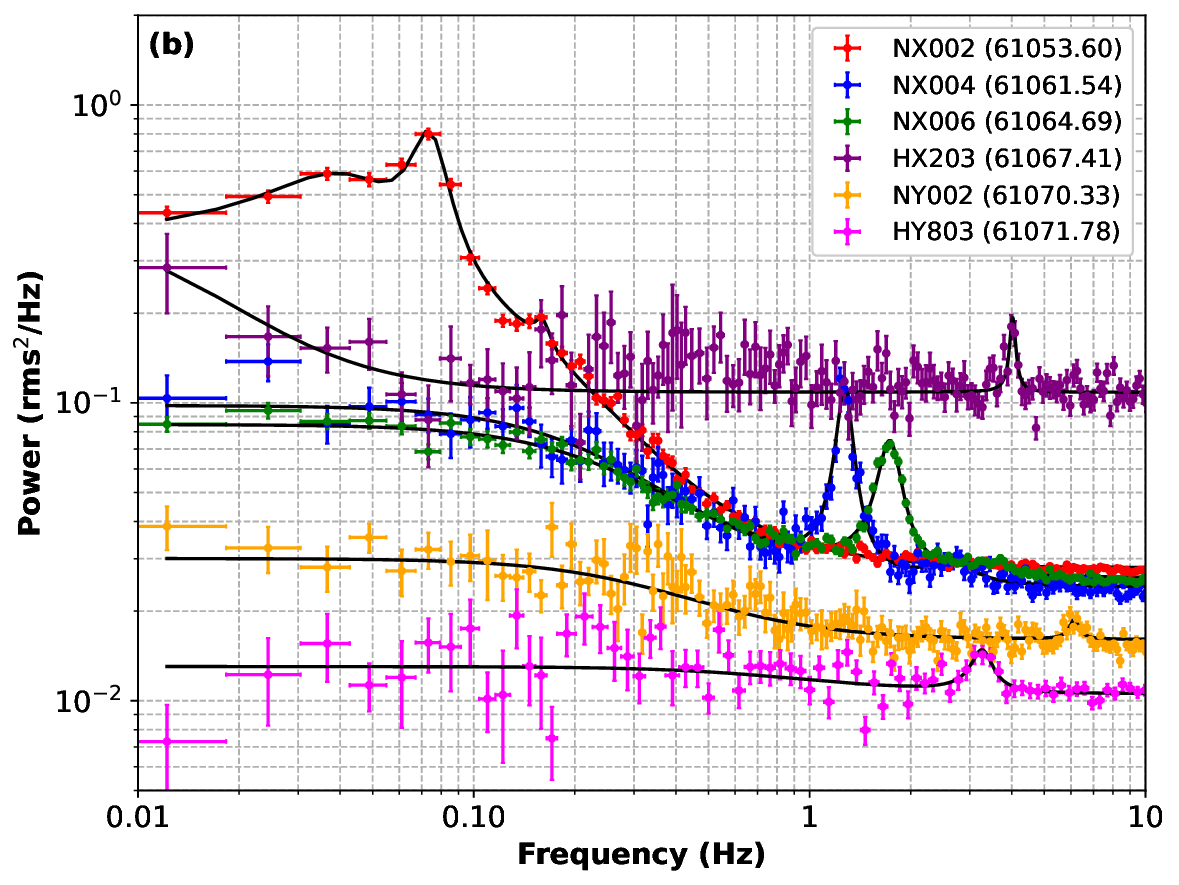}
	\caption{(a) Power Density Spectrum of \textit{NuSTAR} data (ObsID~91201302004), fitted with a combination of 
	three Lorentzian components (one zero-centred and two QPO-centred) and a constant model. (b) PDS of six observations 
	selected from six different evolutionary phases, as marked by points ${\rm a–f}$ in Fig.~\ref{fig_pos}. For clarity, 
	the power in the final observation shown in panel~(b) has been divided by a factor of $10$.} 
\label{fig_pds}
\end{figure*}

The initial phase of the ongoing $2025$–$26$ outburst shows a slow rise with an unusual short-duration ($\sim 3$~days) X-ray flare, 
reaching a peak flux of $\sim 1.4$~Crab (at point~D on 2026 Jan.~29; MJD~61069) in the $2$–$20$~keV band of the GSC. A rebrightening 
episode lasting for $\sim 6$~days, with a peak flux of $\sim 0.8$~Crab (at point~G on 2026 Feb.~16; MJD~61087) in the same energy band 
of {\it MAXI}/GSC, is observed after a gap of $\sim 13$~days.

The higher HR up to point~B (2026 Jan.~14; MJD~61054) indicates that the source was likely in the LHS between points~A
and~B. Owing to an increasing trend in the SXR flux, a declining trend in HR is observed during the period between points~B and~C
(2026 Jan.~14 to 24; MJD~61054–61067). We identify this phase as the HIMS.
The interval between points~C and~E (2026 Jan.~24 to 27; MJD~61067–61069) marks the short X-ray flare, with the CD and DE branches
representing the rising and declining phases of the flare, respectively. During these phases, a rapid rise and fall in the SXR-band flux
is observed with little variation in the HXR band. This behavior implies a sudden inflow of thermally cooler Keplerian matter from the
outer accretion disk, i.e., from the pile-up radius ($X_p$). During this primary flare phase (C–E), the source likely entered
the softer spectral states (HSS or SIMS).
The HR remains at a lower value during the flaring phase, indicating a softening of the spectral nature of the source. Subsequently, 
during the period E–F, the fluxes in both hard and soft bands, as well as the HR, remain nearly constant for $\sim 16$~days. 
We identify this phase of the outburst as a probable SIMS.
The rebrightening of the X-ray flare (F–H) is observed between 2026 Feb.~13 and 19 (MJDs~61084–61090). The HR is further reduced during 
this phase, indicating continued spectral softening. 
A detailed spectral study to confirm the above spectral classification in different phases of the outburst is being carried out and 
will be published elsewhere.

\begin{figure}
\vskip -0.2cm
  \centering
    \includegraphics[angle=0,width=8.8cm,keepaspectratio=true]{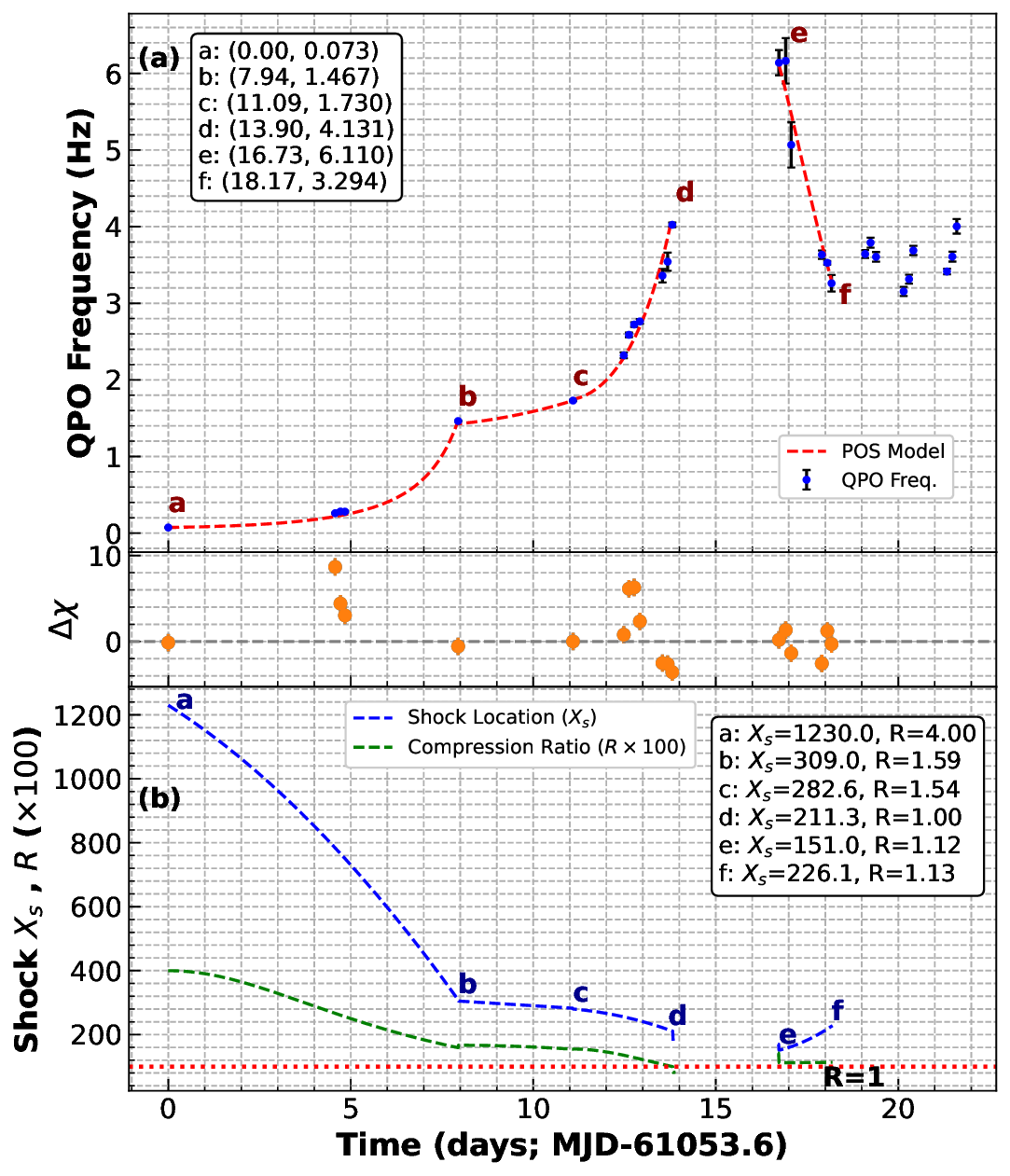}
        \caption{(a) Variations of the observed QPO frequencies with time (in days) during the initial phase of the outburst, along 
        with the best-fitted POS model (dashed curve), are shown. (b) Variations of the shock location (in units of the Schwarzschild 
        radius, $r_s$) and the shock compression ratio ($R$) are presented. The variation of the model-fitted $\Delta\chi$ is shown 
        in the middle panel. Here, points ${\rm a}$–${\rm f}$ mark the start, end, or major transition phases of the evolution. 
        The monotonically increasing and decreasing trends of the QPO frequencies during periods ${\rm a}$–${\rm d}$ and 
        ${\rm e}$–${\rm f}$, respectively, are similar to the rising and declining phases of canonical outbursts of transient BH-XRBs, 
        while the sporadic nature of the QPOs beyond point~${\rm f}$ resembles that observed during the SIMS. }
\label{fig_pos}
\end{figure}

The HID, i.e., the variation of HR with TXR shown in Fig.~\ref{fig_HID}, displays the canonical rising hard and intermediate spectral 
branches characteristic of a classical (type-I) outburst of transient BH-XRBs \citep[see, e.g.,][]{Nandi12}. The vertically rising branch 
A–B indicates the signature of the LHS (rising), while the BC segment corresponds to the HIMS (rising). The flare period C–D–E reflects 
a softer spectral nature of the source. Interestingly, the HR in the declining flare branch (D–E) follows a path similar to that of the 
rising flare branch (C–D), but in the reverse direction. The HR variation during the subsequent nearly constant flux phase (E–F) shows 
a tendency toward the SIMS, which is further supported by the observed QPO evolution. A further decrease in HR is observed during the 
F–H period, which corresponds to the flare rebrightening or second flare phase, indicating that the spectra become even softer during 
this stage of the outburst.

\begin{table}
\addtolength{\tabcolsep}{-4.0pt} 
\vskip -0.3cm
\centering
\caption{Model Fitted QPO Results}
\label{table_qpo}
\renewcommand{\arraystretch}{0.9}  
\begin{tabular}{lccccccc} 
 \hline
Sr.& ObsID & UT  & MJD$_{Avg}$ & $\nu_{QPO}$ & Q & rms(\%) & $\sigma$ \\
\hline
1 & NX2002$^a$& 01/13& 61053.61& $0.073^{+0.003}_{-0.001}$ & $3.21$ & $13.7$ & 7.16 \\
2 & HX0101& 01/18& 61058.18& $0.26^{+0.01}_{-0.02}$ & $8.13$ & $14.9$ & 3.31  \\
3 & HX0102& 01/18& 61058.32& $0.28^{+0.01}_{-0.01}$ & $3.56$ & $18.3$ & 4.88 \\
4 & HX0103& 01/18& 61058.44& $0.28^{+0.01}_{-0.02}$ & $4.40$ & $14.9$ & 2.76 \\
5 & NX2004$^b$& 01/21& 61061.54& $1.46^{+0.01}_{-0.01}$ & $3.93$ & $15.9$ & 23.9 \\
6 & NX2006$^c$& 01/24& 61064.69& $1.73^{+0.01}_{-0.01}$ & $5.28$ & $15.6$ & 22.4 \\
7 & HY0101& 01/26& 61066.08& $2.33^{+0.04}_{-0.05}$ & $7.68$ & $20.0$ & 4.28 \\
8 & HY0102& 01/26& 61066.22& $2.59^{+0.03}_{-0.03}$ & $7.33$ & $21.1$ & 5.86 \\
9 & HY0103& 01/26& 61066.37& $2.72^{+0.03}_{-0.03}$ & $11.1$ & $19.4$ & 4.79 \\
10& HY0104& 01/26& 61066.53& $2.77^{+0.03}_{-0.04}$ & $10.5$ & $18.6$ & 3.65 \\
11& HX0201& 01/27& 61067.14& $3.36^{+0.09}_{-0.09}$ & $4.87$ & $21.4$ & 4.54 \\
12& HX0202& 01/27& 61067.28& $3.53^{+0.12}_{-0.10}$ & $3.68$ & $23.4$ & 5.13 \\
13& HX0203$^d$& 01/27& 61067.41& $3.92^{+0.10}_{-0.06}$ & $12.9$ & $19.7$ & $<4$ \\
14& HY0301& 01/28& 61068.33&                     ---   &                 ---    &                    --- & ---   \\
15& HY0302& 01/28& 61068.49&                     ---   &                 ---    &                    --- & ---   \\
16& HY0303& 01/28& 61068.61&                     ---   &                 ---    &                    --- & ---   \\
17& HY0601& 01/29& 61069.32&                     ---   &                 ---    &                    --- & ---   \\
18& HY0602& 01/29& 61069.47&                     ---   &                 ---    &                    --- & ---   \\
19& HY0603$^e$& 01/29& 61069.60&                     ---   &                 ---    &                    --- & ---   \\
20& NY8002& 01/30& 61070.33& $6.14^{+0.16}_{-0.20}$ & $12.4$ & $14.3$ & $<4$  \\
21& HY0701& 01/30& 61070.52& $6.16^{+0.30}_{-0.28}$ & $5.35$ & $14.3$ & $<3$  \\
22& HY0702& 01/30& 61070.67& $5.07^{+0.30}_{-0.32}$ & $4.00$ & $17.3$ & 3.19  \\
23& HY0801& 01/31& 61071.51& $3.63^{+0.05}_{-0.05}$ & $8.32$ & $20.2$ & 4.85  \\
24& HY0802& 01/31& 61071.66& $3.50^{+0.04}_{-0.04}$ & $11.4$ & $17.8$ & 4.61  \\ 
25& HY0803$^f$& 01/31& 61071.78& $3.26^{+0.11}_{-0.11}$ & $5.54$ & $19.4$ & 3.24  \\
26& HY0901& 02/01& 61072.70& $3.64^{+0.05}_{-0.06}$ & $6.66$ & $21.1$ & 4.38  \\
27& HY0902& 02/01& 61072.84& $3.79^{+0.07}_{-0.07}$ & $7.11$ & $20.2$ & 3.78  \\
28& HY0903& 02/01& 61072.99& $3.61^{+0.06}_{-0.06}$ & $7.20$ & $21.1$ & 4.08  \\
29& HY1001& 02/02& 61073.75& $3.16^{+0.06}_{-0.06}$ & $6.13$ & $20.2$ & 4.81  \\
30& HY1002& 02/02& 61073.90& $3.32^{+0.06}_{-0.06}$ & $5.74$ & $23.1$ & 4.55  \\
31& HY1101& 02/03& 61074.93& $3.42^{+0.04}_{-0.04}$ & $9.66$ & $17.9$ & 4.13  \\
32& HY1102& 02/04& 61075.08& $3.61^{+0.07}_{-0.07}$ & $9.34$ & $16.7$ & $<3$  \\
33& HY1103& 02/04& 61075.21& $4.00^{+0.09}_{-0.08}$ & $8.98$ & $20.8$ & 3.01  \\
34& HY1301& 02/06& 61077.10& $4.59^{+0.12}_{-0.12}$ & $8.43$ & $16.9$ & $<3$  \\ 
35& HY1302& 02/06& 61077.23& $4.46^{+0.76}_{-0.32}$ & $3.10$ & $20.5$ & $<3$  \\   
36& HY1401& 02/07& 61078.30& $4.25^{+0.23}_{-0.18}$ & $6.01$ & $19.4$ & $<3$  \\    
37& HY1402& 02/07& 61078.44& $4.33^{+0.36}_{-0.34}$ & $2.78$ & $24.3$ & 3.27  \\ 
38& HY1403& 02/07& 61078.56& $3.86^{+0.10}_{-0.11}$ & $9.23$ & $19.4$ & $<3$  \\    
39& HY1901& 02/08& 61079.07& $4.75^{+0.22}_{-0.18}$ & $5.07$ & $19.8$ & 3.20  \\    
40& HY1903& 02/08& 61079.34& $5.02^{+0.11}_{-0.09}$ & $8.51$ & $7.07$ & 3.00  \\ 
41& NZ1002& 02/11& 61082.76& $4.67^{+0.05}_{-0.04}$ & $8.22$ & $7.67$ & 7.33  \\
42& NZ1004& 02/14& 61085.17& $5.81^{+0.13}_{-0.12}$ & $12.6$ & $4.51$ & $<4$  \\
\hline
\end{tabular}
\leftline{NX=9120130, NY=9120230, NZ=9120231, HX=P08140470,}
\leftline{HY=P08140480 are pre-fixes of the NuSTAR and HXMT} 
\leftline{observation IDs. UT in dd/mm format and $\nu_{QPO}$ in Hz.}
\leftline{Q ($=\nu_{QPO}/FWHM$), \% of rms, and significance ($\sigma$) define}
\leftline{nature of the QPOs. Note that some of the HXMT} 
\leftline{observations show $\sigma \sim 3$ due to less statistics.}
\end{table}

\subsection{Evolution of Low Frequency QPOs}

A strong signature of low-frequency QPOs is observed in the PDS of each {\it NuSTAR} and {\it Insight}-HXMT data, except during 
the softer phase of the outburst (2026 Jan 28–29; MJD 61068.33–61069.60), which covers part of the primary X-ray flare. The details of the 
Lorentzian model fitted QPO frequency ($\nu$), Q-value ($\nu$/FWHM), fractional rms (\% rms = $100~\sqrt{\pi \times \text{power} \times 
\text{FWHM}/2}$), and significance ($\sigma$, defined as the ratio of the norm to its negative error of the Lorentzian 
model fitted QPOs) are listed in Table~\ref{table_qpo}.

A monotonic evolution of the QPO frequencies is observed during the initial phase of the outburst. In the very first {\it NuSTAR} 
observation on 2026 Jan 13 (MJD 61053.61), a type-C QPO at 0.073~Hz is detected. In the follow-up observations with {\it NuSTAR} 
and {\it Insight}-HXMT, the QPO frequencies increase monotonically up to $\sim 4$~Hz at point C in Fig.~\ref{fig_lc}(a) (2026 Jan 27; 
MJD 61067.41). After that, no QPOs are detected for the next $\sim 2$~days, which includes the peak of the first flare. 
The QPO reappears in the declining phase of the flare on 2026 Jan 30 (MJD 61070.33). Subsequently, it shows a monotonically 
decreasing trend over the next $\sim 1.5$~days, until 2026 Jan 31 (MJD 61071.78). 
During this shorter evolution phase, the QPO frequency rapidly decreases from $6.14$~Hz to $3.26$~Hz, a behavior generally observed 
during the canonical declining phase of classical outbursts in transient BH-XRBs. This type of monotonically decreasing trend in the 
QPO frequency immediately after an increasing trend, or during the rising phase of an outburst, is uncommon. 
However, a similar evolution was observed in Swift~J1727.8-1613 during its recent discovery outburst \citep{D25}.

\begin{figure*}
  \centering
    \includegraphics[angle=0,width=8.8cm,keepaspectratio=true]{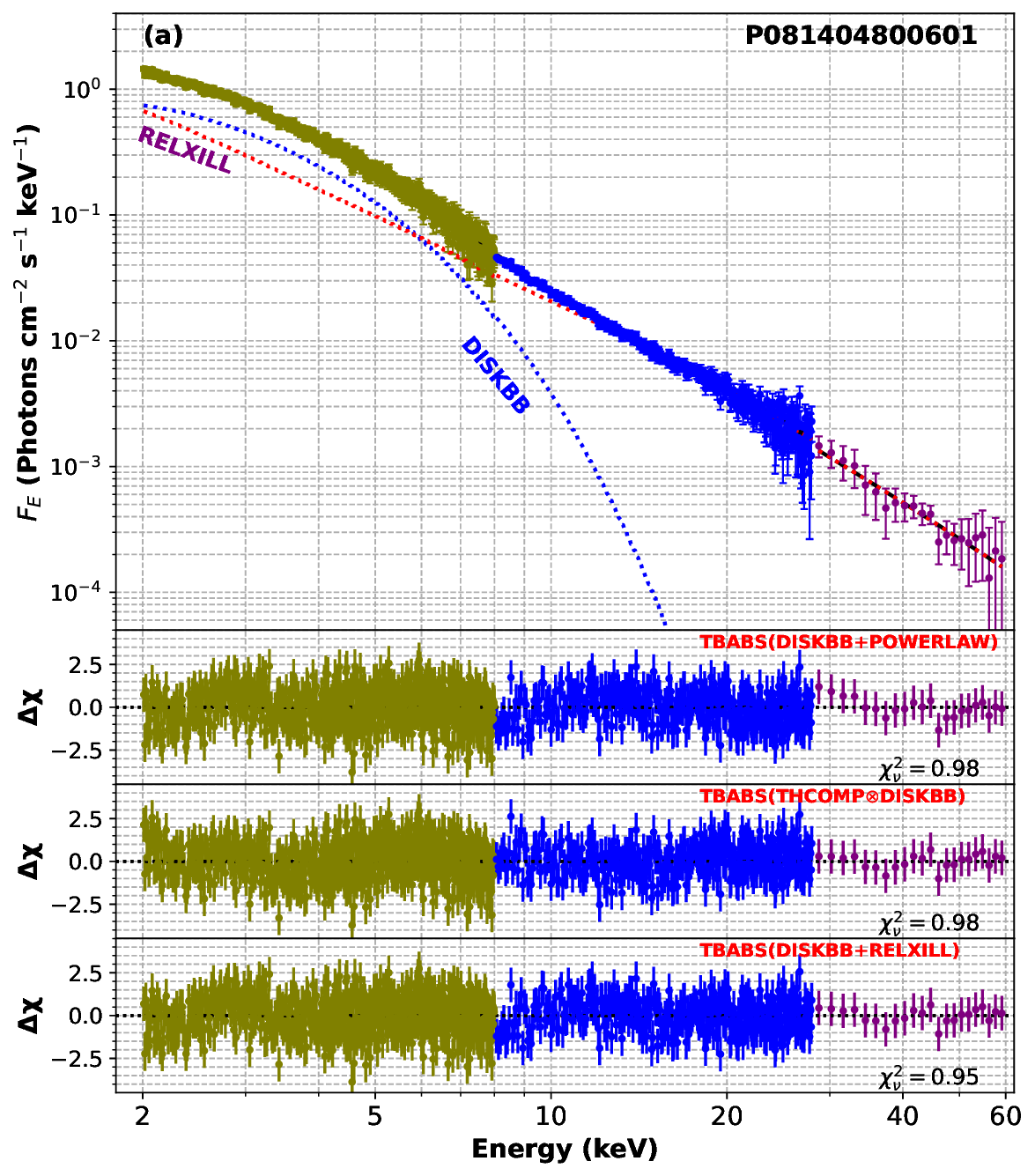}\hskip 0.2cm
    \includegraphics[angle=0,width=8.8cm,keepaspectratio=true]{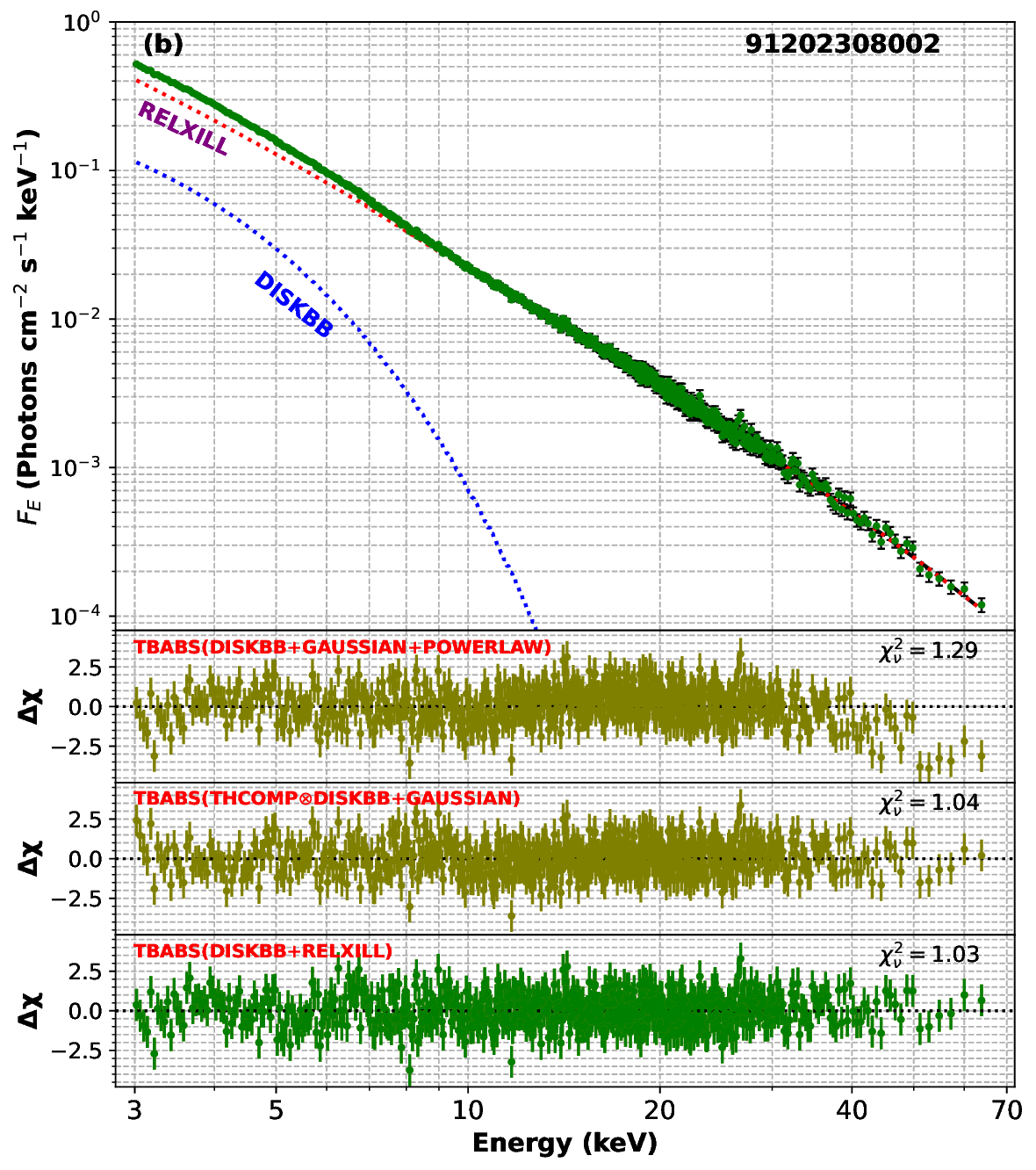}
	\caption{Broadband spectra of (a) \textit{Insight-HXMT} (ObsID P081404800601; MJD 61069.32; SPEC1) and (b) \textit{NuSTAR}/FPMA 
    (ObsID 91202308002; MJD 61070.33; SPEC2), close to the first flare peak, are fitted with different combinations of spectral models 
    M1, M2 and M3 (see \S 2.4). The lower panels show the residuals, expressed as the variations of the model-fitted 
    $\Delta\chi$, for the corresponding models (marked in the insets). See text for details.} 
\label{fig_spec}
\end{figure*}

\begin{table*}
\addtolength{\tabcolsep}{-4.0pt} 
\vskip -0.3cm
\centering
\caption{Model Fitted Spectral Parameters with Different Components of Respective Models}
\label{table_spec}
\renewcommand{\arraystretch}{0.9}  
\begin{tabular}{lccccccccccc}
 \hline\hline
\multicolumn{3}{c}{\textsc{$^1$Tbabs(diskbb + powerlaw)}} & / & \multicolumn{5}{c}{\textsc{$^4$Tbabs(diskbb + gaussian + powerlaw)}} \\
\hline
ObsID & $kT_{\rm in}$  & N$_{\rm DBB}$ & $\Gamma$ & N$_{\rm PL}$ & $\chi^2/DOF$ &  &  &  \\
\hline
HY601$^1$& $1.23^{+0.01}_{-0.01}$ & $264^{+19}_{-19}$ & $2.41^{+0.02}_{-0.02}$ & $5.43^{+0.39}_{-0.38}$ &1037/1056 &  &  &   \\
NY002$^4$& $1.24^{+0.03}_{-0.02}$ & $53^{+6.4}_{-6.3}$ & $2.65^{+0.01}_{-0.01}$ & $10.1^{+0.28}_{-0.27}$ & 808/626  & & &  \\
 \hline\hline
\multicolumn{3}{c}{\textsc{$^2$Tbabs(thcomp$\otimes$diskbb)}} & / & \multicolumn{5}{c}{\textsc{$^5$Tbabs(thcomp$\otimes$diskbb + gaussian)}} & \\
\hline
ObsID & $\Gamma_\tau$  & $kT_e$ & Cov$_{frac}$ & $kT_{\rm in}$ & N$_{\rm DBB}$ &  $\chi^2/DOF$ & F$_{THCOMP}$ & $F_{DBB}$ & FR    \\
\hline
HY601$^2$& $2.49^{+0.03}_{-0.03}$ & $<30$ & $0.48^{+0.04}_{-0.04}$ & $0.99^{+0.02}_{-0.02}$ &  $1110^{+99}_{-85}$ & 1030/1056 & 8.1 & 5.8 & 1.4  \\
NY002$^5$& $2.50^{+0.03}_{-0.03}$ & $41^{+9}_{-8}$ & $0.72^{+0.07}_{-0.06}$ & $0.95^{+0.02}_{-0.02}$ & $840^{+64}_{-55}$ & 651/625 & 7.5  & 3.4 & 2.2 \\
 \hline\hline
\multicolumn{9}{c}{$^3$\textsc{Tbabs(diskbb+relxill)}}  \\
\hline
ObsID & $kT_{\rm in}$ & N$_{\rm DBB}$ & $a_k$ & $i$ & $\Gamma$ & $\log\xi$ & A$_{Fe}$ & E$_{cut}$ & refl$_{frac}$ & N$_{RELXILL}$ & $\chi^2/DOF$   \\
\hline
HY601$^3$& $1.23^{+0.01}_{-0.01}$ & $293^{+11}_{-10}$ & $0.998^{+p}_{-0.002}$ & $64.8^{+2.0}_{-2.1}$ & $2.46^{+0.05}_{-0.04}$ & 2.99$^{+0.07}_{-0.13}$ &  $<0.65$ & $200^f$ & 3.78$^f$ & $0.33^{+0.01}_{-0.01}$ & 1004/1053 \\
NY002$^3$& $1.17^{+0.02}_{-0.03}$ & $90^{+5.1}_{-4.7}$ & $0.998^{+p}_{-0.001}$ & $68.4^{+1.6}_{-1.8}$ & $2.40^{+0.02}_{-0.03}$ & 3.69$^{+0.24}_{-0.42}$ & $<5.0$ &  $65^{+3.8}_{-4.2}$ & 0.38$^{+0.02}_{-0.04}$ & $0.45^{+0.01}_{-0.01}$ & 642/624 \\
\hline
\end{tabular}
\leftline{HY=P0814048006 and NY=91202308 are the pre-fixes of the Insight-HXMT (H) and NuSTAR (N) observation IDs.}
\leftline{$^f$ Unable to constrain error limits, model parameters are kept frozen at their best fitted values.}
\leftline{Fixed values of TBABS: $N_H=7\times10^{21}$~{\rm cm}$^{-2}$; RELXILL: Index1, Index2$=9$, R$_{br}=15$, R$=-1$, and R$_{out}=400$ are used.}
\leftline{Units of $kT_{in}$, $kT_e$, E$_{cut}$ in keV; R$_{br}$ in ISCO radius; R$_{out}$ in Gravitational radius $r_g=GM/C^2$.}
\leftline{Thermal DISKBB and nonthermal THCOMP fluxes F$_{\rm DBB}$, and F$_{\rm THCOMP}$ are in $10^{-9}~{\rm erg}~{\rm cm}^{-2}~{\rm s}^{-1}$.}
\leftline{FR is the ratio of the nonthermal to thermal fluxes.} 
\leftline{Note, model fitted parameter errors are the average values of the $1\sigma$ $\pm$ errors and spin upper limit is pegged to the hard bound. }
\end{table*}

After that, a non-evolving, sporadic nature of the QPOs is observed for the next $\sim 13$~days, until 2026 Jan 14 (MJD 61085.17). 
This type of sporadic QPO behavior is commonly seen during the SIMS of transient BH-XRB outbursts. For instance, similar non-evolving, 
sporadic QPOs were observed during the SIMS of the recent outburst of Swift~J1727.8$-$1613 \citep{D25} and the 2010–11 outburst of 
GX~339$-$4 \citep{Nandi12}.

We further study the monotonic evolution (both increasing and decreasing) of the QPO frequencies using the propagating oscillatory 
shock (POS) model. This approach allows us to understand the properties of the shock wave, whose oscillation is to be considered the origin of 
the observed low-frequency QPOs \citep[see, e.g.,][]{C08,D10,D25}. In Fig.~\ref{fig_pos}, we show the evolution of the QPOs fitted 
with the POS model. The shock is observed to move inward (from $1230~r_s$ to $211~r_s$, where $r_s$ is the Schwarzschild radius) 
during the monotonically increasing phase of the QPOs, marked by points ${\rm a-d}$. Conversely, during the short monotonically 
decreasing phase of the QPO frequency (region between ${\rm e-f}$), the shock recedes from $151~r_s$ to $226~r_s$.

Based on the POS model--fitted shock parameters, such as the shock strength and velocity, we classify the evolution of the QPO frequency 
into four stages: $(i)$ a--b, $(ii)$ b--c, $(iii)$ c--d, and $(iv)$ e--f. Faster shock motion is observed during phases $(i)$ and $(iv)$. 
In these four phases, the shock moves inward or outward with velocities of $\sim 2$, $0.2$, $0.25$, and $0.8$~m~s$^{-1}$, respectively. 
The strong shock (compression ratio $R=4$) at point $a$ weakens to $R\simeq 1$ by the last day of the increasing QPO phase at point $d$. 
During the receding phase $(iv)$, the shock remains weak, with only a very slight increase in strength.

During the rising phase, due to the slowing down of the shock movement between points $b$ and $c$ ($\sim 3$~days), only a small increase in the 
QPO frequency ($1.46$--$1.73$~Hz) is observed. After that, over a roughly similar time interval between points $c$ and $d$, a sharp 
increase in the QPO frequency from $1.73$ to $4$~Hz is observed. During this rapid QPO increasing phase, although the shock velocity 
remains nearly the same, the rapid weakening of the shock strength ($R$ reduced from $\sim 1.54$ to $1$) is likely to be the primary 
cause of the sharp rise in the QPO frequency.

\subsection{Spectral Modeling: On the nature of Spin}

Broadband spectra using LE ($2$--$8$~keV), ME ($8$--$29$~keV), and HE ($29$--$60$~keV) data from {\it Insight}-HXMT and {\it NuSTAR} 
are used for the spectral analysis. The primary goal is to understand the nature of the source during the unconventional flare phase. 
We selected the flare-peak data from {\it Insight}-HXMT (ObsID P081404800601 on MJD 61069.32; hereafter referred to as SPEC1) with 
an exposure of $\sim 2$~ks and the nearest {\it NuSTAR} observation (ObsID 91202308002 on MJD 61070.33; hereafter referred to as SPEC2) 
with an exposure of $\sim 12$~ks, as marked by the yellow and cyan shaded regions in Fig.~\ref{fig_lc}. In Fig.~\ref{fig_spec}, we show 
both spectra fitted with three different models (see \S2.4). A strong signature of relativistic reflection is observed in both 
observations.

To understand the overall spectral nature of the source during the short flare-peak phase, we first fitted both spectra with the 
phenomenological model M1, comprising a thermal \texttt{disk blackbody} plus a nonthermal \texttt{powerlaw} component. 
Although model M1 fits SPEC1 reasonably well, it fails to adequately describe the broader energy band ($3$--$70$~keV) higher 
resolution {\it NuSTAR} spectrum, particularly above $\sim 50$~keV. A broad Fe emission line at 
$\sim 6.5$~keV is also evident in the NuSTAR spectrum. Higher disk temperatures ($T_{\rm in} > 1.2$~keV) and steep photon indices 
($\Gamma > 2.4$) in both observations indicate that the source was in a softer (HSS and SIMS) spectral state. 

We then replaced the nonthermal \texttt{powerlaw} component with the physical thermal Comptonization model \texttt{thcomp}. This 
provides a better understanding of the `hot' corona, which is responsible for producing the nonthermal hard photons. Model 
M2 is used to fit both spectra. But an additional \texttt{gaussian} component for the Fe emission line 
is required to achieve a satisfactory fit in SPEC2. It is to be noted that the physical \texttt{thcomp} model is quite successful in 
fitting the nonthermal continuum, including the feature above $50$~keV. An increase in corona temperature ($kT_e$), and 
the covering fraction of the thermal seed photons in SPEC2 indicates that the corona becomes relatively hotter, with a possible increase 
in its size and optical depth. Furthermore, we note that the spectrum becomes slightly harder even within the softer state, 
as indicated by an increase in the flux ratio (FR), defined as the ratio of nonthermal to thermal flux. Note that due to the 
lower spectral resolution and lower exposure of HXMT, no prominent Fe line is observed in SPEC1.

Finally, to estimate intrinsic source parameters such as spin and inclination angle, and to understand the spectral nature from 
a physical perspective, both spectra were fitted using the relativistic reflection model \texttt{relxill}. For SPEC2, this model 
did not require an additional Gaussian Fe line profile. The lower disk temperature ($T_{in}$) and photon index ($\Gamma$) indicate 
that the source moved toward a slightly harder spectral state during SPEC2. The higher high-energy cutoff ($E_{\rm cut}$) suggests 
that the corona size may have been smaller during SPEC1. A higher relativistic reflection fraction in the SPEC1 is also consistent 
with it. The source spin is estimated to be $a_k = 0.998^{+p}_{-0.002}$ and $0.998^{+p}_{-0.001}$, at inclination angles 
$i=(65.2^{+2.4}_{-2.1})^\circ$ and $(68.4^{+1.6}_{-1.8})^\circ$ for SPEC1 and SPEC2, respectively, from model M3. 
The high spin values indicate that the source is a maximally rotating Kerr black hole. The estimated spin and inclination 
angle of the source are consistent with earlier results reported by \citet{El-Batal16} and \citet{Casares09}.

\section{Discussion and Concluding Remarks}

The initial phase of the 2025–26 outburst of the Galactic transient BH GS~1354-64 is studied using data from {\it MAXI}/GSC, 
{\it Insight-HXMT}, and {\it NuSTAR}. The source exhibited an unconventional outburst profile, although it began with a common 
trend of slow-rise. The outburst included a short X-ray flare of $\sim 3$~days, followed by a rebrightening after $\sim 13$~days. 
The maximum fluxes during the peaks of the primary and secondary flares were observed to be $1.4$~Crab and $0.8$~Crab, respectively. 
Note that the outburst is still ongoing at a flux level similar to that of the plateau phase (E--F).

The on-demand data from {\it MAXI}/GSC were used to study the outburst profile of the source. The daily-averaged GSC fluxes in the 
soft ($2-6$~keV; SXR), hard ($6-20$~keV; HXR), and total ($2-20$~keV; TXR) X-ray bands, along with their ratio defined as the hardness 
ratio (HR = HXR/SXR), allow us to characterize the spectral behavior of the source during the 2025–26 outburst. From the evolution of 
the fluxes in different energy bands, it is evident that the outburst was primarily driven by thermal soft photons, as the TXR 
closely followed the trend of the SXR, while the HXR flux remained roughly constant throughout the outburst.

The points A–H in Fig.~\ref{fig_lc} mark important evolutionary phases of the outburst profile and HR. In the early phase of the 
outburst, between points A and B, the HR remained roughly constant at higher values. The vertical branch in the HID (Fig.~\ref{fig_HID}) 
during this phase indicates that the source was in the LHS. The phase between points~B and C is defined as the HIMS, as a decreasing 
trend in the HR with an increase in the SXR is observed. During the primary flare period (between points C–D–E), a nonconventional 
evolution of the HR is observed. The declining flare branch (D–E) follows the same returning path as the rising flare branch (C–D). 
The source was found to be in its softest phase on the peak day (point D) of the primary flare, as indicated by the lowest HR. 
After this, the source remained roughly constant in both X-ray intensity and HR between points E and F. Starting from point F, the 
HR decreased and remained at lower values during the second flare (between points F–G–H). Although the timescales and physical 
processes associated with GS~1354-64 (a BH-XRB) and pulsars (neutron stars) are different, the HR and flux plateau phase (E--F) 
between the two flares appears similar to the bridge between the on- and off-pulse phases of a pulsar.

The physical origin of the two intense, short-duration flares may be attributed to a sudden increase in the supply of high-viscosity
Keplerian matter into the disk from a temporary reservoir at the pile-up radius $X_p$, leading to a short-term enhancement in the SXR
rate. A rapid decline in TXR is observed immediately after the first flare. However, the plateau phase between the two flares 
indicates a fresh supply of matter from $X_p$ following the initial decay of the first flare, which likely triggered the second flare 
of the outburst. From the evolution of the outburst profile, HR, and HID, it appears that the $2025$--$2026$ outburst of GS~1354$-$64 does 
not correspond to a canonical type-I or type-II outburst. A more detailed spectral study of the entire $2025$--$2026$ outburst of GS~1354$-$64 
will be carried out and presented elsewhere.

A strong signature of low-frequency QPOs is observed in both \textit{Insight-HXMT} and \textit{NuSTAR} observations throughout 
the outburst, except during the initial $\sim 2$~days of the primary X-ray flare. Prior to their disappearance, the QPO frequencies 
exhibit a monotonic increase from $73$~mHz to $4$~Hz (between points $a$–$d$ in Fig.~\ref{fig_lc}b). After reappearing, the QPOs show 
a decreasing trend (from $6.14$~Hz to $3.26$~Hz) over $\sim 1.5$~days (between points $e$–$f$), which is typically seen during the 
declining phase of transient BH candidate outbursts. Subsequently, non-evolving sporadic QPOs are observed during the plateau epoch (E–F) of 
the outburst, a behavior commonly associated with the SIMS of transient BH-XRBs.

To investigate the shock-wave dynamics during the QPO evolution, we fitted the observed frequencies with the POS model, in which 
LFQPOs are considered to originate from oscillations of the shock. Important phases of the POS model fitted QPO evolution is marked 
by points $a$-$f$. The model fit indicates that the shock moves inward toward the BH (from $1230$ to $211~r_s$) between points $a$ and 
$d$, and recedes outward (from $151$ to $226~r_s$) between points $e$ and $f$. A higher shock velocity is observed during the $a$–$b$ 
and $e$–$f$ phases. Additionally, the shock strength gradually weakens during the QPO increasing phase $a$–$d$, reaching a minimum 
compression ratio of $R=1$ at point $d$.

The observed QPO trends are broadly consistent with those commonly seen during the rising and declining harder spectral states 
(HS and HIMS) of transient BH candidates, where inward shock propagation (corresponding to increasing QPO frequency) and outward propagation 
(corresponding to decreasing QPO frequency) are typically observed. However, an outward shock motion associated with a monotonic 
decrease in the QPO frequency immediately following the end of the monotonic increase in the QPO evolution is uncommon.
A similar short-duration QPO frequency decrease was recently reported in Swift~J1727.8-1613 for $\sim 1.2$~days within an 
overall increasing trend \citep{D25}. From a physical perspective, an outward-moving shock may occur when viscosity at the outer edge 
of the disk (i.e., at $X_p$) decreases or is temporarily turned off, thereby reducing the supply of high-viscosity Keplerian matter, 
which constitutes the primary source of thermal soft X-ray photons. 
The roughly constant QPO frequency between $b$ and $c$, after the initial rise, occurs due to the slow inward movement of 
the shock wave. This behavior is likely associated with a roughly constant or slowly increasing thermal Keplerian component, as 
seen in the SXR band. A similar feature was observed in the 1999 outburst of XTE~J1859+226 \citep{Nandi18} and in the recent 
2023--24 outburst of Swift~J1727.8-1613 \citep{D25}.

To understand the nature of the source during the softest phase of the outburst, i.e., during the primary flare epoch, we performed 
broadband spectral analysis using one \textit{Insight-HXMT} observation (on the peak day at point D; ObsID P081404800601) and one 
\textit{NuSTAR} observation (ObsID 91202311002), marked as yellow and cyan shades in Fig.~\ref{fig_lc}, respectively. We employed 
three model sets combining phenomenological and physical models. The phenomenological \texttt{diskbb+powerlaw} model provides a basic 
picture of the thermal and nonthermal components, but does not reveal the physical mechanisms behind their production. A prominent 
broad Fe emission line at $\sim 6.5$~keV is observed in the NuSTAR spectrum (SPEC2). The higher disk temperature ($T_{\rm in}>1.2$~keV) 
and photon index ($\Gamma > 2.4$) indicate that the source was in the softer spectral states (HSS or SIMS) during the primary flare. 

To investigate the properties of the `hot' Compton cloud (corona), we replaced the \texttt{powerlaw} with the physical thermal 
Comptonization model (\texttt{thcomp}), defining model M2. However, this model along with a \texttt{gaussian} 
component for the prominent Fe emission line, is found to adequately fit the nonthermal continuum, including the deviation 
in the higher-energy band ($>50$~keV) of SPEC2. The higher coronal temperature ($kT_e$) and increased covering fraction in the 
second observation suggest an increase in both size and optical depth of the corona. The lower disk temperature ($kT_{\rm in}$) 
also indicates a comparatively harder spectral state. Furthermore, a decrease in thermal \texttt{diskbb} flux relative 
to the nonthermal \texttt{thcomp} flux suggests that the source became harder in the later NuSTAR observation, which 
explains the reappearance of the QPO during the NuSTAR observation period. Out of the total $\sim 40$~ks NuSTAR 
observation period, the QPO at $\sim 6.14$~Hz is significantly detected only after $\sim 25$~ks. 

Finally, spectral fitting with the relativistic \texttt{relxill} model provides a more detailed physical picture of the source. 
The \texttt{relxill+diskbb} model fits both spectra well, and no additional \texttt{powerlaw} component is required for SPEC2. 
The decrease in the thermal disk temperature ($T_{\rm in}$) and photon index ($\Gamma$) in SPEC2 confirms that the source 
transitioned to a slightly harder state compared to SPEC1. Since SPEC1 corresponds to the softest phase, a higher cutoff energy 
($E_{\rm cut}$) is observed. These results imply that the corona may have been smaller during SPEC1 of \textit{Insight-HXMT}. 
The higher relativistic reflection fraction in SPEC1 is consistent with this scenario.

We also estimated intrinsic parameters, namely the spin ($a_k$) and inclination angle ($i$), from the spectral fits using model 
M3. For SPEC1, we obtained $a_k = 0.998^{+p}_{-0.002}$ and $i = (65.2^{+2.4}_{-2.1})^\circ$. For SPEC2, we measured 
$a_k=0.998^{+p}_{-0.001}$ and $i = (68.4^{+1.6}_{-1.8})^\circ$. Combining these results, we predict the spin and inclination of 
the source to be $a_k \sim 0.998$ and $i \sim 63^\circ-70^\circ$. These measurements are consistent with earlier reports by 
\citet{El-Batal16} and \citet{Casares09}.

Recent \textit{IXPE} observations from 2026 Feb 7–9 detected a significant polarization degree of $\sim 4\%$ at a polarization 
angle of $\sim -1^\circ$ in the $2-8$~keV energy band \citep{Ravi26}. The spin and inclination derived from simultaneous spectral modeling of 
\textit{IXPE} and \textit{Insight-HXMT} data are consistent with the results presented here. A detailed spectro-polarimetric study 
has been performed and will be published elsewhere.

A brief summary of our findings in this {\it paper} is as follows:

\begin{enumerate}[i)]
    \item GS 1354-64 exhibited an unconventional outburst profile, characterized by two short-duration X-ray flares. The peak fluxes 
    of the two flares were observed to be $1.4$ and $0.8$~Crab in the $2-20$~keV band of {\it MAXI}/GSC.
    \item The soft X-ray {\it MAXI}/GSC flux in $2-6$~keV followed a trend similar to the total flux in $2-20$~keV, while the hard 
    X-ray flux in $6-20$~keV remained roughly constant. This indicates that thermal photons from the high-viscosity Keplerian disk 
    were the primary component controlling the evolution of the spectral and temporal properties of the source.
    \item The HID resembled the canonical rising phase of a transient BH candidate outburst. However, during 
    the onset and decay phases of the primary flare, the hardness ratio (HR) retraced the same path, which is quite uncommon.
    \item Strong signatures of LFQPOs were observed throughout the outburst (except for $\sim 2$~days during 
    the primary X-ray flare) where {\it Insight}-HXMT and {\it NuSTAR} data were available. Monotonic evolution of the QPO 
    frequencies was observed up to the declining phase of the primary flare. Thereafter, non-evolving sporadic QPOs were observed 
    for $\sim 13$~days between the two flares.
    \item Fitting the QPO frequency evolution with the propagating oscillatory shock (POS) model allowed us to probe the nature of 
    the shock, which is considered the origin of the LFQPOs according to the shock oscillation model. A stronger shock moved inward 
    from $1230~r_s$ to $211~r_s$ and weakened during the monotonically increasing phase of the QPO frequency between 2026 Jan 1–27 
    (MJD 61053.61–61067.41). During the short flare decay phase (MJD 61070.33–61071.78), a weaker receding shock moved outward 
    from $151~r_s$ to $226~r_s$.
    \item Detailed spectral analysis of two observations (one each from HXMT and NuSTAR) using both physical and phenomenological 
    models indicates that the source was in the softer (HSS and/or SIMS) states during the primary flare.
    \item The source became slightly harder (within the softer spectral phase) in the later NuSTAR observation during the decay 
    phase of the primary flare. This explains the reappearance of the LFQPO in the late epoch of the NuSTAR observation on 
    2026 Jan 30.
    \item Relativistic modeling with \texttt{relxill} estimates the spin and inclination angle as $a_k \sim 0.998$ and 
    $i \sim 63^\circ$–$70^\circ$, consistent with previous reports in the literature.
\end{enumerate}

\section*{Acknowledgements}

We are thankful to the anonymous referee and scientific Editor for kind suggestions to improve the quality of the paper.
This work made use of on-demand data from \textit{MAXI}, a Japanese Experiment Module on the International Space Station, and 
archival data from {\it Insight}-HXMT, a satellite mission of the China National Space Administration (CNSA) and the Chinese 
Academy of Sciences (CAS), as well as \textit{NuSTAR}, a NASA observatory. The authors sincerely thank the MAXI team (RIKEN, JAXA) 
for providing on-demand data products and the HXMT team for providing the latest CALDB file (v2.08) prior to its 
official release.
DD acknowledges the visiting research grant from National Tsing Hua University, Taiwan (NSTC 114-2811-M-007-084). 
HKC and SS acknowledge support from the National Science and Technology Council (NSTC) of Taiwan under grant NSTC 114-2112-M-007-042.
AN thanks GH, SAG; DD; PDMSA; and the Director, URSC, for encouragement and continuous support in carrying out this research. 
We also thank the instrument teams for processing the data and providing the necessary software tools for the analysis.

\end{document}